\documentclass[twocolumn]{aastex701}
\accepted{\today}
\begin{document}

\title{PSF-like Alpha-Particle Events in LSST Images}

\author[0000-0001-6013-1131,gname='Guillem',sname='Megias Homar']{Guillem Megias Homar}
\affiliation{California Institute of Technology, Pasadena, CA 91125, USA}
\email[show]{gmegias@caltech.edu}
\correspondingauthor{Guillem Megias Homar}

\author[0000-0002-9601-345X,gname='Craig S.',sname='Lage']{Craig~S.~Lage}
\affiliation{Physics Department, University of California, One Shields Avenue, Davis, CA 95616, USA}
\email{cslage@ucdavis.edu}

\author[0000-0002-8357-3984,gname='Pierre-François',sname='Léget']{Pierre-Fran\c{c}ois~L\'eget}
\affiliation{Department of Astrophysical Sciences, Princeton University, Princeton, NJ 08544, USA}
\email{leget@astro.princeton.edu}

\author[0000-0003-4833-9137,gname='Steven M.',sname='Kahn']{Steven~M.~Kahn}
\affiliation{Physics Department,  University of California, 366 Physics North, MC 7300 Berkeley, CA 94720, USA}
\email{stevkahn@berkeley.edu}

\author[0000-0003-0347-1724,gname='Christopher W.',sname='Stubbs']{Christopher~W.~Stubbs}
\affiliation{Department of Astronomy, Center for Astrophysics, Harvard University, 60 Garden St., Cambridge, MA 02138, USA}
\affiliation{Center for Astrophysics, Harvard \& Smithsonian, 60 Garden Street, Cambridge, MA 02138}
\affiliation{Department of Physics, Harvard University, 17 Oxford St., Cambridge MA 02138, USA}
\email{stubbs@g.harvard.edu}

\author[orcid=0000-0001-5390-8563,sname='Kulkarni']{S. R. Kulkarni}
\affiliation{California Institute of Technology, Pasadena, CA 91125, USA}
\email{srk@astro.caltech.edu}  

\author[0000-0001-8708-251X,gname='Ian S.',sname='Sullivan']{Ian~S.~Sullivan}
\affiliation{University of Washington, Dept.\ of Astronomy, Box 351580, Seattle, WA 98195, USA}
\email{sullii@uw.edu}

\author[0000-0003-2759-5764,gname='James F.',sname='Bosch']{James~F.~Bosch}
\affiliation{Department of Astrophysical Sciences, Princeton University, Princeton, NJ 08544, USA}
\email{jbosch@astro.princeton.edu}

\author[0000-0001-9376-3135,gname='Eli S.',sname='Rykoff']{Eli~S.~Rykoff}
\affiliation{Kavli Institute for Particle Astrophysics and Cosmology, SLAC National Accelerator Laboratory, 2575 Sand Hill Rd., Menlo Park, CA 94025, USA}
\email{erykoff@stanford.edu}

\begin{abstract}

Rare $\alpha$-particle–induced charge clusters appear in LSST images as compact, PSF-like sources with a median FWHM of 0\farcs95 and median ellipticity consistent with zero, closely resembling unresolved astrophysical point sources. These events are detected in both dark and science exposures at a rate of approximately $10^{-12}\ \mathrm{pixel}^{-1}\ \mathrm{s}^{-1}$. Their collected charge and morphology are consistent with energy deposition from $\sim$5 MeV $\alpha$-particles in silicon CCDs, and their spatial distribution across the focal plane suggests a localized material origin, plausibly associated with trace radioactive contamination in the cryostat aluminum. Despite their deceptive appearance, we demonstrate that a simple broadness statistic based on fourth-order moments cleanly separates these events from stellar PSFs, enabling efficient rejection in coadded images and real-time alert streams. Such charge clusters do not impose an intrinsic bright-end contamination floor for Rubin transient searches, as genuine fast astrophysical events would exhibit characteristically different morphological signatures.

\end{abstract}

%% PASP uses Unified Astronomy Thesaurus (UAT) concepts:

\section{Introduction}

The NSF–DOE Vera C. Rubin Observatory Legacy Survey of Space and Time \citep[LSST;][]{Ivezic2019} has been widely described as a ``transient discovery machine,'' yet its capability to probe variability on sub-minute timescales remains largely unexplored. In LSST images, millisecond transients would manifest with structured morphological signatures \citep{Megias2023}. However, the main challenge lies in distinguishing genuine astrophysical events from non-astrophysical ``twinkles'', including satellite glints \citep{loeb2024,tyson2024,tanaka2025}, cosmic rays, and other particle–induced detector artifacts. The scientific rewards are broad: optical counterparts to GRBs (and FRBs), sub-minute flares from red dwarfs, and, perhaps, entirely new phenomena.

In this work, we investigate a class of compact, PSF-like sources in LSST images produced by heavily ionizing particles in silicon CCDs. Such localized charge depositions have been extensively studied in other contexts, particularly in dark matter direct detection experiments, where radioactive contamination and particle backgrounds are critical systematics. It is well established that $\alpha$-particles deposit sufficient charge density to satisfy the plasma condition (Debye length is small compared with charge-column dimensions), producing large, round ionization clusters in fully depleted silicon. $\alpha$-particles lose energy predominantly via ionization, generating a dense column of electron–hole pairs within a short stopping distance \citep{Estrada_2011}. 

\begin{figure*}[t!]
    \centering
    \includegraphics[width=\linewidth]{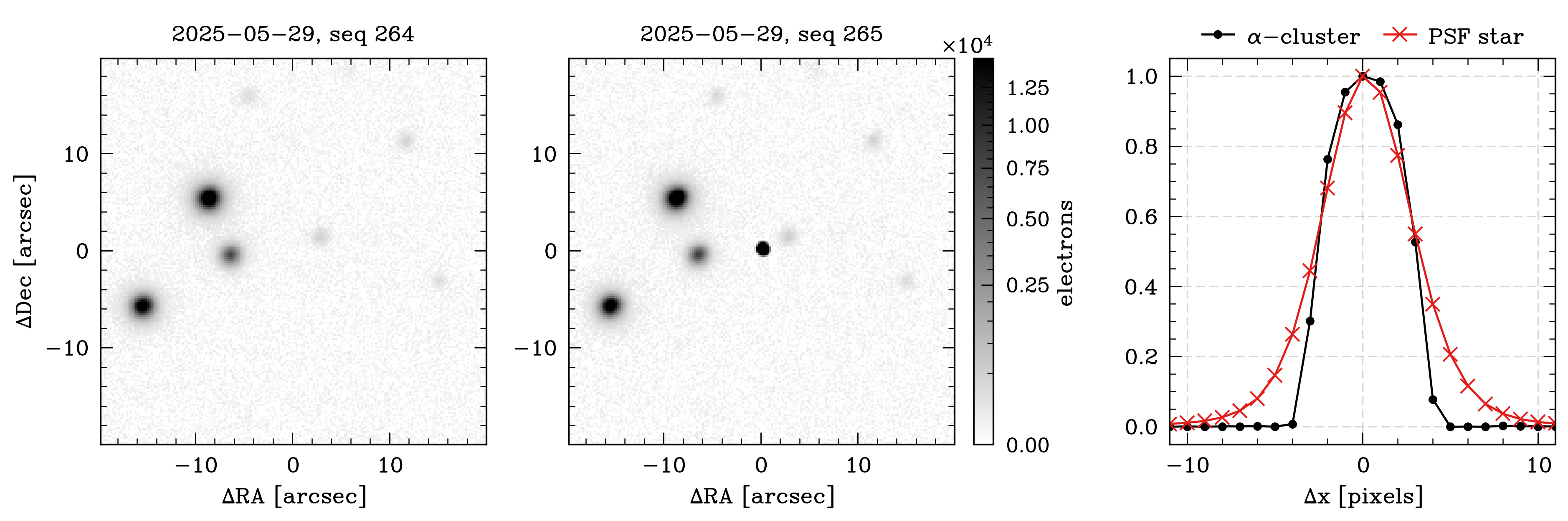}
    \caption{(Left, Center) Example of a PSF-like $\alpha$-cluster detected in only one of two consecutive 30-s exposures obtained on 2025-05-29 (sequences 264 and 265; detector 63) at $\mathrm{RA}=327.163^\circ$, $\mathrm{Dec}=-13.209^\circ$. Pixel values are shown in units of electron–hole pairs. (Right) One-dimensional profile across the source, comparing the $\alpha$-cluster to a median PSF star, normalized to the same peak flux. The profiles closely match in the core but diverge at larger radii.}
    \label{fig1:twinkle}
\end{figure*}

The Rubin Observatory LSST Camera \citep[LSSTCam;][]{Camera2025} is a 3.2-gigapixel CCD mosaic composed of 10~$\mu$m pixels and 100~$\mu$m-thick fully depleted sensors. This architecture differs from systems such as DECam (15~$\mu$m pixels, $\sim$250~$\mu$m thickness) and DAMIC (15~$\mu$m pixels, $\sim$675~$\mu$m thickness), in which $\alpha$-particle-induced charge clusters have been extensively characterized \citep{Estrada_2011,Aguilar_Arevalo_2015}. Differences in pixel scale and depletion depth affect charge transport, diffusion, and the morphology of high-ionization clusters. 

\citet{Grosson_2023} previously investigated unusually ``fat'' cosmic-ray tracks in Rubin sensors, hypothesizing that slow cosmic-ray protons generating charge clouds of order $10^4$ electrons broaden during drift due to charge repulsion. However, that study did not consider $\alpha$-particles, which generate larger and denser charge clouds ($\sim10^6$ electrons) within a short stopping range in silicon. Given their very short stopping ranges (tens of microns), these particles must originate locally from radioactive contamination within the instrument rather than from cosmic sources. Although rare in LSST images, we show that, coincidentally, these clusters have characteristic sizes comparable to the atmospheric Point Spread Function (PSF) under typical seeing conditions, closely mimicking unresolved astrophysical sources and complicating their identification in LSST alerts, difference imaging and time-domain searches.

The LSST image processing pipelines \citep{pipelines2025}, whose cosmic-ray rejection algorithms trace back to SDSS \citep{Lupton_2002}, detect cosmic rays by identifying sharp pixel-level outliers relative to PSF-weighted directional thresholds, merging adjacent flagged pixels into footprints, and interpolating over contaminated regions. While effective for elongated or sharply peaked cosmic-ray tracks, this approach fails for clusters that appear round and PSF-like.

Here, we present a systematic characterization of $\alpha$-particle-induced clusters (``$\alpha$-clusters'') in LSST images. We begin by describing the LSSTCam data and the selection methodology used to identify $\alpha$-clusters in both dark and science exposures (\S\ref{data}). We then quantify their occurrence rates, spatial distribution across the focal plane, and total collected charge distribution, using these diagnostics to constrain their likely instrumental origin (\S\ref{results}). Finally, we examine their morphology and show that a simple broadness-based statistical metric distinguishes them from genuine astrophysical point sources, discussing the implications for Rubin’s time-domain searches (\S\ref{discussion}).

\section{Data}\label{data}

$\alpha$-particle clusters were first spotted serendipitously during a search for fast optical transients in pairs of consecutive 30-s exposures acquired during commissioning and engineering tests, while the telescope remained fixed on the same field. We analyzed 3,020 exposure pairs (``snaps''), totaling 6,040 exposures. All images were processed through the standard LSST calibration pipelines, including Instrument Signature Removal (ISR) \citep{Plazas_2025}, cosmic ray rejection, image calibration and source detection. Non-repeating source candidates were identified by catalog matching and difference imaging. Forced photometry was then performed at each candidate location in both frames to confirm the non-repeating nature of the signal. Figure~\ref{fig1:twinkle} shows an example of an $\alpha$-cluster detected in only one of two consecutive 30-s LSST exposures.

In parallel, we also analyzed 2,092 30-s darks obtained at the observatory site (Cerro Pachón, Chile) between May 2025 and January 2026. These data were processed through ISR, cosmic-ray rejection and source detection. The detected sources included some elongated tracks, single-pixel events, and the compact PSF-like clusters studied here. We isolated the latter by measuring their second moments. In both datasets, we imposed a minimum integrated charge of $5\times10^5$ electrons, avoiding poor image matches and subtractions.

\begin{figure*}[t]
    \centering
    \includegraphics[width=\linewidth]{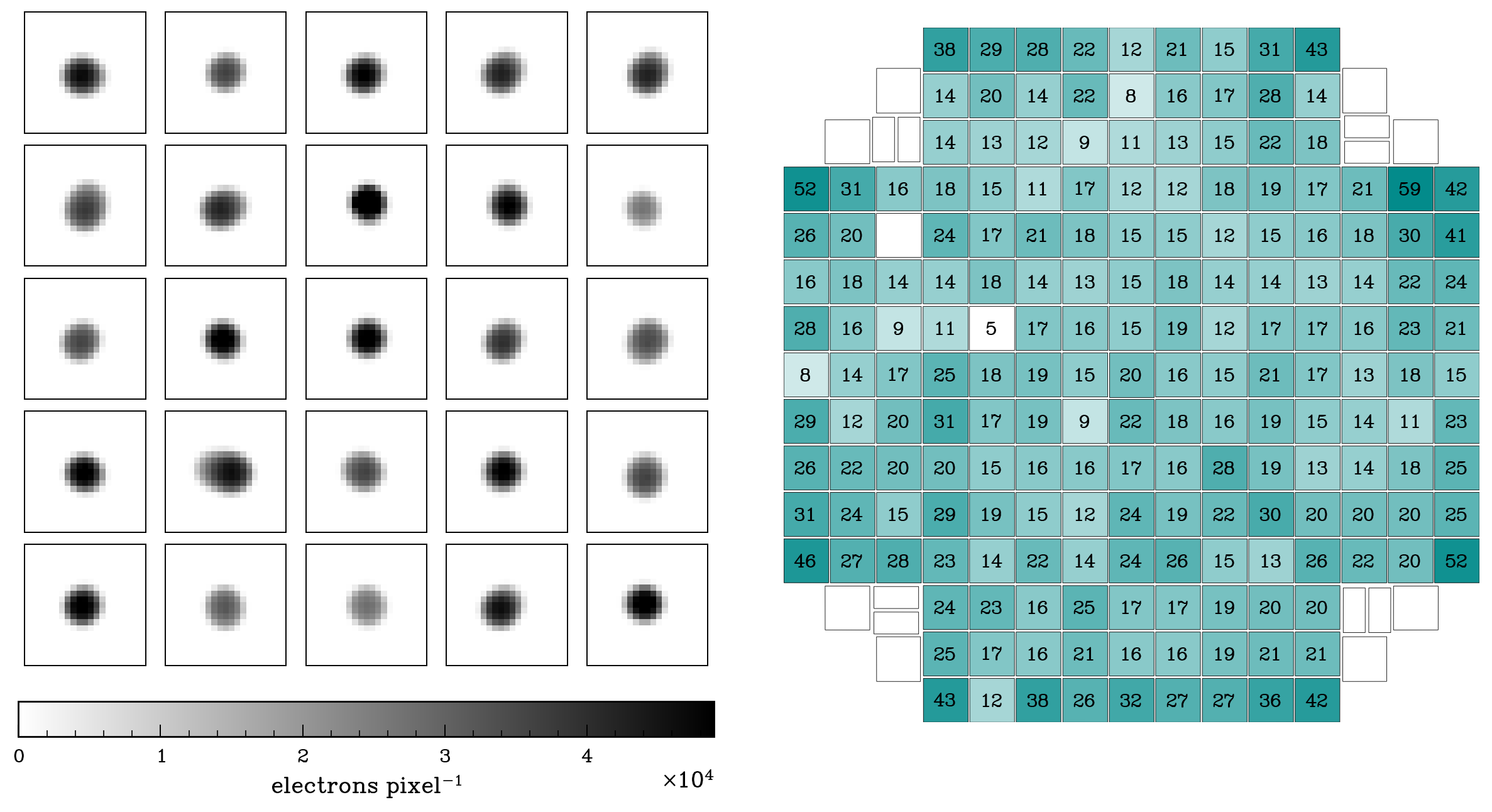}
    \caption{(Left) Sample of twenty-five $\alpha$-clusters identified in dark exposures, illustrating the range of observed morphologies. While many clusters appear nearly circular, others exhibit mild trailing. (Right) Spatial distribution of all detected clusters across the focal plane, combining science and dark samples. Although the total number of clusters is limited, clusters are more frequently observed toward the edges of the focal plane. The blank detector corresponds to a non-operational CCD. The corner rafts are not included in the study.}
    \label{fig:twinkle_zoo}
\end{figure*}

\section{Results}\label{results}

Across the 6,040 science exposures (3,020 snap pairs), we identified 2,411 $\alpha$-clusters. In the 2,092 darks, we detected 1,414 $\alpha$-clusters. These correspond to an occurrence rate of $(4.15 \pm 0.08)\times10^{-12}$ pixel$^{-1}$ s$^{-1}$ in science images, and $(7 \pm 0.2)\times 10^{-12}$ pixel$^{-1}$ s$^{-1}$ in darks, where uncertainties reflect $\sqrt{N}$ counting statistics. The difference between these rates is highly significant statistically, and likely reflects higher detection efficiency in darks, as $\alpha$-clusters in science images can overlap with astronomical sources or lie in crowded regions, causing them to fail deblending or quality-selection cuts and therefore be excluded from the sample. Overall, the combined rate is of order $(4$--$7)\times10^{-12}$ pixel$^{-1}$ s$^{-1}$, corresponding to approximately 0.4--0.6 clusters per 30-s LSST exposure (roughly one every two exposures), confirming that these events are rare but not negligible.

A representative sample of these $\alpha$-clusters, together with their spatial distribution across the focal plane for the combined datasets, is shown in Figure~\ref{fig:twinkle_zoo}. Most clusters appear compact and nearly round, closely resembling the PSF, though a few exhibit mild trailing. The focal-plane distribution reveals a greater incidence toward the edge of the focal plane. There does not appear to be a positional dependence of the mildly trailed clusters.

\begin{figure}[h!]
    \centering
    \includegraphics[width=\linewidth]{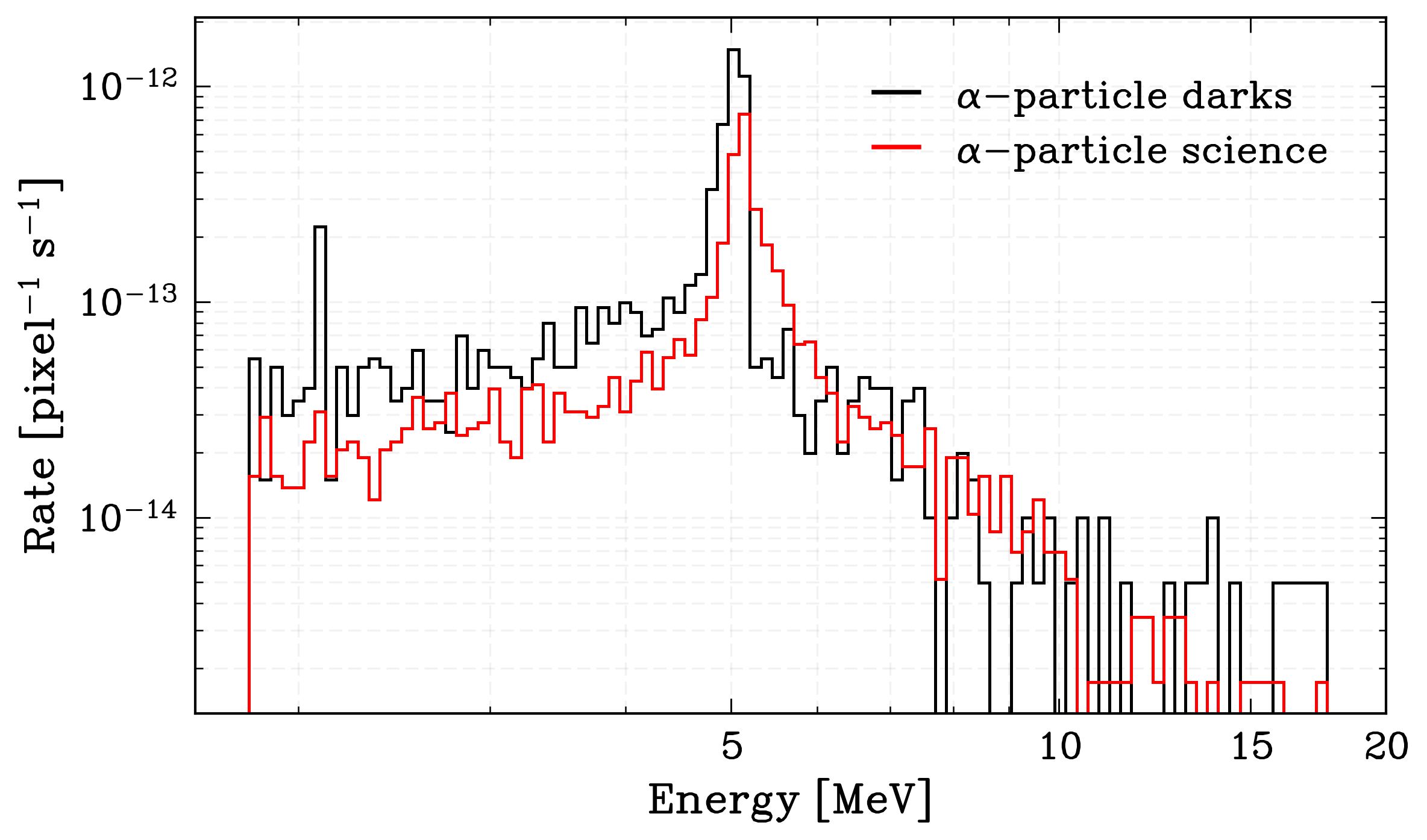}
    \caption{Energy spectra of the identified $\alpha$-clusters in dark and science exposures. The y-axis gives the event rate in each energy bin, expressed in units of pixel$^{-1}$\,s$^{-1}$ and computed from the number of counts per bin. The science sample shows a broader peak, likely reflecting the larger flux uncertainty caused by sky-background contributions.}
    \label{fig:distribution}
\end{figure}

In science images, the collected charge is measured within a fixed circular aperture of 12-pixel in diameter, while in darks it is summed over the detected footprint. Using the known mean ionization energy of 3.6 eV per electron–hole pair in silicon \citep{McKay1953}, we convert the measured charge to deposited energy in MeV. The resulting energy distribution is shown in Figure~\ref{fig:distribution}, with both science and dark samples peaking near $\sim5$ MeV ($\sim1.4\times10^6$ electrons).

\begin{figure*}[th!]
    \centering
    \includegraphics[width=0.65\linewidth]{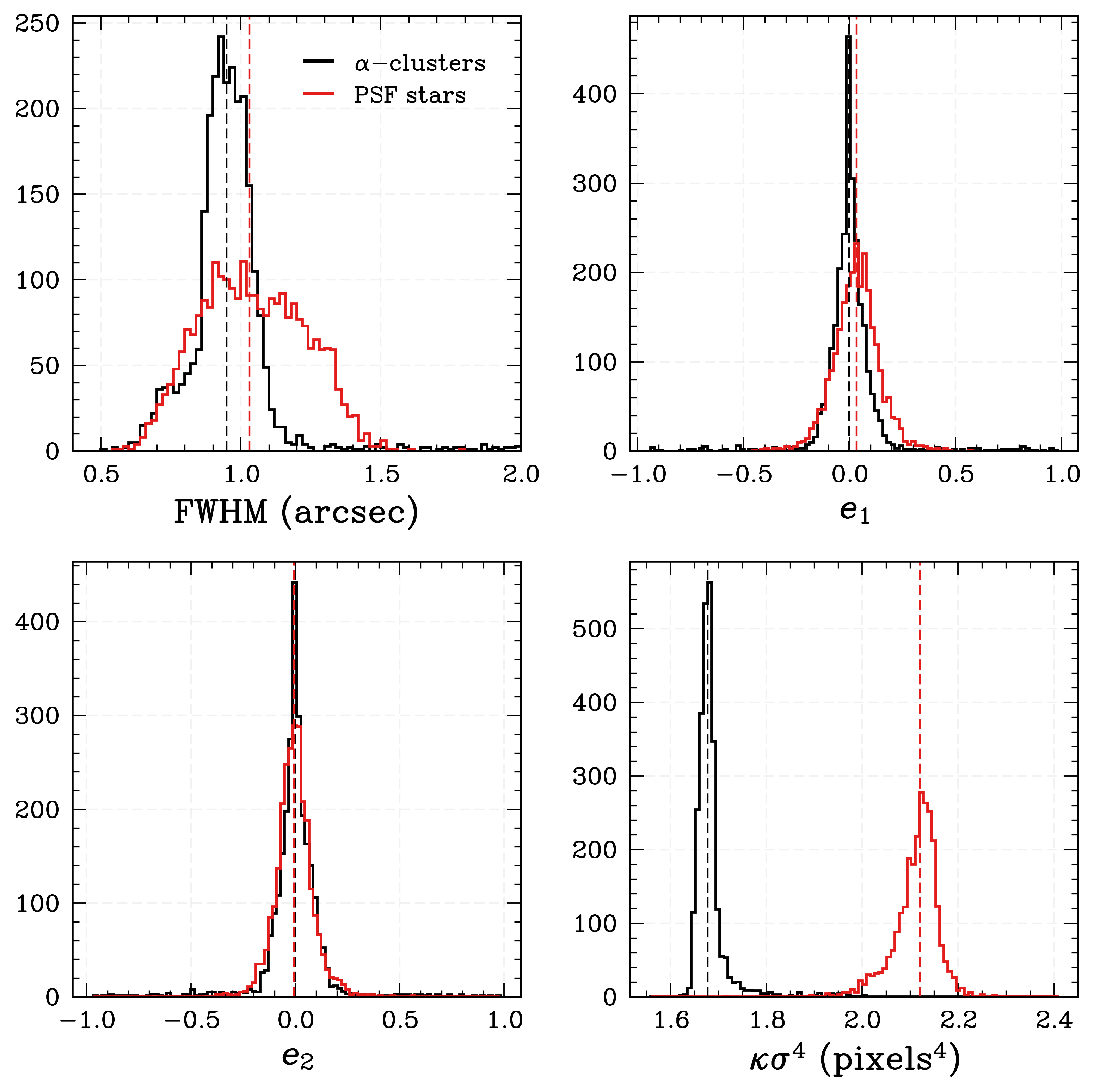}
    \caption{Second-order moments, FWHM, and broadness of the detected $\alpha$-clusters in science images. Black histograms show the $\alpha$-cluster population, while red histograms show median PSF stars on the corresponding detectors for comparison. Here $I_{xx}$ and $I_{yy}$ are the adaptive second central moments along the detector axes, with $T = I_{xx}+I_{yy}$, and the ellipticity components are defined as $e_1 = (I_{xx}-I_{yy})/T$ and $e_2 = 2I_{xy}/T$. The dashed vertical lines mark the median values of each distribution. The ellipticity distribution is centered near zero, indicating predominantly round morphologies, with a small number of large excursions corresponding to rare trailed events. The median $\alpha$-clusters FWHM is $\sim0\farcs95$, corresponding to $T \approx 8.2$ pixels$^2$, whereas median stellar FWHM is $\sim1\farcs02$. The broadness metric (proportional to the fourth central moment) highlights the suppressed wings of the $\alpha$-clusters relative to stellar PSFs.}
    \label{fig:morphology}
\end{figure*}

Morphologically, $\alpha$-clusters are compact and nearly round. Figure~\ref{fig:morphology} shows the distributions of FWHM, ellipticity, and a higher-order broadness statistic, $\kappa\sigma^4$, defined as
\begin{equation}
    \kappa\sigma^4 = M_{40} + 2M_{22} + M_{04}
\end{equation}
where $M_{ij}$ are adaptive Gaussian-weighted central moments \citep{Hirata2003, Mandelbaum2005}. Here, $\sigma^2$ denotes the adaptive second-moment size and $\kappa$ is the kurtosis. For a surface-brightness distribution with variance $\sigma^2$, $\kappa$ measures the relative weight of the wings compared to the core. The quantity $\kappa\sigma^4$ (\textit{broadness}) therefore represents the unnormalized fourth central moment, directly encoding how much flux resides at large radii.

For comparison, Figure~\ref{fig:morphology} also shows the distribution of median PSF stars on the corresponding detectors. Although clusters resemble the PSF in second-moment properties, their higher-order statistic reveals systematically sharper profiles.

\section{Discussion}\label{discussion}

We begin by considering the energy distribution of the identified clusters. In both darks and science exposures, the distribution peaks near $\sim5\,$MeV. As seen in Figure~\ref{fig:distribution}, the peak in the dark sample is narrower than in the science images. We attribute this broadening in science exposures to additional flux uncertainty from sky background contributions, as fluxes were measured using a fixed 12-pixel circular aperture.

To assess whether these signals are consistent with $\alpha$-particle interactions, we compare their observed sizes with expectations from semi-empirical plasma-effect models. \citet{Estrada_2011} derived an expression for the lateral size ($\sigma$) of plasma-dominated clusters,
\begin{equation}
    \sigma^2 \propto \Bigg[\frac{E}{a_{\mathrm{eff}} E^{b_{\mathrm{eff}}} + c_{\mathrm{eff}}}\Bigg]^2
\end{equation}
where $E$ is the $\alpha$-particle energy in MeV and the other coefficients are derived from fits to the $\alpha$ range in silicon, $R(E) = aE^b + c$, as first derived by \cite{Gobeli1956}.

Using \citet{Estrada_2011}'s fitted parameters ($a_{\mathrm{eff}} = 0.85$, $b_{\mathrm{eff}} = 0.84$, $c_{\mathrm{eff}} = 0$) and scaling to LSSTCam’s 10$\,\mu$m pixels (instead of Estrada's 15$\,\mu$m), we obtain an expected $\sigma_{5 \mathrm{MeV}} \approx 2.28\,$pixels, corresponding to $\sim1\farcs06$ FWHM. This is close to the median PSF FWHM observed in science images ($0\farcs95$, as seen in Figure~\ref{fig:morphology}). Overall, the observed charge distribution and morphology are consistent with $\sim5\,$MeV $\alpha$-particle deposition in silicon.

Having established this consistency, we now consider the likely origin of $\alpha$-clusters. A $\sim5$ MeV $\alpha$-particle has a stopping range in silicon of $R(E = 5\,\mathrm{MeV})\sim 3\,\mu \mathrm{m}$ \citep{Gobeli1956}, implying that the observed plasma-dominated morphology requires particles to enter from the backside and deposit their energy close to the entrance surface. The greater frequency of occurrence toward the edges of the focal plane further suggests a localized material source, plausibly associated with structural components near or in the cryostat.

Recycled aluminum (Al) typically contains trace contamination from the natural U and Th decay chains at the ppm to sub-ppm level \citep{Saisho1988DeterminationOT}. Although the use of virgin Al was considered, the LSSTCam cryostat ultimately employed recycled material, implying the presence of multiple $\alpha$-emitting isotopes with characteristic energies in the 4–6 MeV range. In particular, isotopes in the $^{238}$U and $^{235}$U (Actinium) series, including $^{231}$Pa, can produce $\alpha$ emissions consistent with the observed $\sim5\,$MeV cluster energies. Welding rods used in Al fabrication can also contain $^{232}$Th, introducing localized contamination at weld joints and potentially contributing to the greater incidence near the edges.

For instance, consider a cryostat with an inner radius of $\sim64\,$cm, like that of the LSSTCam. Because the stopping range of a $\sim5\,$MeV $\alpha$-particle in Al is only of order a few microns, and the detectable particles must emerge from the aluminum into the silicon, only the top $\sim5\,\mu$m of the inner Al surface can plausibly contribute to the observed signal. That thin surface layer alone contains $\sim 50$–$60$ g of Al. On the assumption of a U contamination level of 0.1 ppm, this corresponds to $\sim5\times10^{-6}\,$g of U and $\sim6\times10^3$ decay-chain decays per day in that thin surface layer. Because the $\alpha$ emission is expected to be approximately isotropic, only about half of these particles are emitted into the inward-facing hemisphere, giving an effective rate of order $\sim3\times10^3$ potentially detectable $\alpha$-particles per day before accounting for additional geometric acceptance. By contrast, the measured $\alpha$-cluster rate corresponds to $\sim10^3$ clusters per day across the focal plane. This order-of-magnitude agreement supports a radiogenic origin for the PSF-like clusters, arising from trace U/Th contamination in structural Al components of the cryostat.

While these clusters are rare, they occur frequently enough to contaminate coadded images, alert streams, and difference images. In second-moment quantities (FWHM and ellipticity), they closely resemble PSF stars, with a median FWHM of $0\farcs95$ (compared to $1\farcs02$ for stars) and ellipticity consistent with zero. However, unlike atmospheric PSFs, they lack extended wings and are therefore intrinsically sharper.

This difference is revealed by higher-order moments. Because $\alpha$-clusters are more centrally concentrated and have suppressed wings relative to stellar PSFs, they exhibit systematically lower \textit{broadness}, $\kappa\sigma^4$, values. As shown in Figure~\ref{fig:morphology}, this statistic cleanly separates $\alpha$-clusters from stellar sources on the same detector. For the empirical cut $\kappa\sigma^4 \gtrsim 1.9\,\mathrm{pixels}^4$, only $1.26\%$ of the $\alpha$-cluster sample would pass the cut, corresponding to a rejection efficiency of $98.74\%$. Conversely, only $0.27\%$ of stellar sources fall below this threshold, corresponding to a stellar-PSF retention rate of $99.73\%$. Because this sample spans multiple nights and observing conditions, these values indicate that broadness is a robust, high-purity discriminator. Incorporating this metric into the LSST alert stream would provide a simple and effective improvement in artifact rejection.

This naturally raises the question of whether such rejection criteria could inadvertently suppress real sub-minute Fast Optical Bursts. While $\alpha$-clusters are temporally non-repeating, they are intrinsically sharper and lack atmospheric PSF structure. In contrast, even very short astrophysical transients propagate through the atmosphere and are therefore imprinted by atmospheric seeing. As discussed in \citet{Megias2023}, the detailed morphology depends on burst duration relative to atmospheric turbulence timescales: longer bursts average over many turbulence realizations and approach a seeing-broadened stellar PSF, while millisecond bursts sample a more instantaneous, speckled atmospheric PSF. In both cases, however, the source remains distributed over the atmospheric seeing disk, including extended or speckled wings, rather than producing the compact morphology of an $\alpha$-cluster. We therefore do not expect a broadness-based cut of the kind considered here to significantly suppress genuine Fast Optical Bursts. Nevertheless, careful threshold optimization will remain important in an operational alert stream, especially for events where the atmospheric wings or speckled structure are detected at low signal-to-noise.

\section{Conclusion}
We have characterized $\alpha$-particle–induced clusters in LSST images that appear nearly PSF-like, with a median FWHM of $0\farcs95$ and ellipticity centered near zero. Their spatial distribution, with a greater incidence toward the focal-plane edges, indicates a localized radiogenic origin within the instrument, plausibly associated with trace U/Th decay-chain contamination in structural aluminum components of the cryostat. Although rare ($(4–7)\times10^{-12}$ events pixel$^{-1}$ s$^{-1}$), $\alpha$-clusters occur frequently enough to contaminate coadds, alert streams, and difference images in fast transient searches. Crucially, we show that they are systematically sharper than the stellar PSFs and can be robustly distinguished using fourth-order moment statistics, specifically the $\kappa\sigma^4$ broadness metric. We plan to incorporate such rejection strategies into the LSST processing pipelines.

\begin{acknowledgments}
We thank the Rubin Data Management Team, without whom this work would not have been possible, including R. H. Lupton, Y. AlSayyad, L. MacArthur, N. Lust, L. Kelvin, M. Fisher-Levine, B. Kalmbach, C. Waters, M. Gower, and F. Moolekamp. We are also grateful to F. Harrison, A. Roodman, T. Tyson, P. Antilogus, A. Rasmussen, and J. R. Peterson for valuable discussions throughout this project. GMH thanks the Buttermilk Society for numerous insightful discussions.

This material is based upon work supported in part by the National Science Foundation through Cooperative Agreement AST-1258333 and Cooperative Support Agreement AST-1202910 managed by the Association of Universities for Research in Astronomy (AURA), and the Department of Energy under Contract No. DE-AC02-76SF00515 with the SLAC National Accelerator Laboratory managed by Stanford University. Additional Rubin Observatory funding comes from private donations, grants to universities, and in-kind support from LSST-DA Institutional Members.

Lage gratefully acknowledges support from DOE grant DE-SC0009999.
\end{acknowledgments}

\facilities{Rubin:Simonyi(LSSTCam)}

\bibliography{PASPsample701}{}
\bibliographystyle{aasjournalv7}



\end{document}